# $MnO_2$-gated Nanoplatforms with Targeted Controlled Drug Release and Contrast-Enhanced MRI Properties: from 2D Cell Culture to 3D Biomimetic Hydrogels

Yupeng Shi, Flavien Guenneau, Xiaolin Wang#, Christophe Hélary, Thibaud Coradin✉

Sorbonne Université, CNRS, Collège de France, Laboratoire de Chimie de la Matière Condensée de Paris, 75005 Paris, France.

#Current address: State Key Laboratory of Quality Research in Chinese Medicine and School of Pharmacy, Macau University of Science and Technology, Macau.

✉ Corresponding author: Thibaud Coradin E-mail: thibaud.coradin@sorbonne-universite.fr; Tel: +33-1-44274018

## Abstract

Multifunctional nanomaterials combining diagnosis and therapeutic properties have attracted a considerable attention in cancer research. Yet some important challenges are still to be faced, including an optimal coupling between these two types of properties that would be effective within complex biological tissues. To address these points, we have prepared novel nanoplatforms associating controlled drug delivery of doxorubicin and Magnetic Resonance Imaging (MRI) contrast-enhancement that exhibit high specificity towards cancer cells compared to normal cells and evaluated them both in 2D cultures and within 3D tissue-like biomimetic matrices.

Methods: Nanoplatforms were prepared from hollow silica nanoparticles coated with $MnO_2$ nanosheets and conjugated with the AS1411 aptamer as a targeting agent. They were fully characterized from a chemical and structural point of view as well as for drug release and MRI signal enhancement. Standard two-dimensional monolayer cultures were performed using HeLa and Normal Human Dermal Fibroblasts (NHDF) cells to testify targeting and cytotoxicity. Cellularized type I collagen-based hydrogels were also used to study nanoparticles behavior in 3D biomimetic environments.

Results: The as-established nanoplatforms can enter HeLa cells, leading to the dissociation of the $MnO_2$ nanosheets into $Mn^{2+}$ that enhanced $T_1$ magnetic resonance signals and concomitantly release doxorubicin, both effects being markedly more significant than in the presence of NHDFs. Moreover, particles functionality and specificity were preserved when the cells were immobilized within type I collagen-based fibrillar hydrogels.

Conclusion: The use of $MnO_2$ nanosheets as glutathione-sensitive coatings of drug loaded nanoparticles together with surface conjugation with a targeting aptamer offers an effective strategy to obtain efficient and specific nanotheranostic systems for cancer research, both in 2D and 3D. The here-described tissue-like models should be easy to implement and could constitute an interesting intermediate validation step for newly-developed theranostic nanoparticles before *in vivo* evaluation.

## Introduction

Despite the rapid improvement of modern medicine, the early diagnosis and therapy of cancer is still a challenge. The continuous development of nanotechnology and the emergence of targeted treatments provide an inequivalent opportunity in this area. The past decade has witnessed the engineering of many theranostic nanosystems, where the integration of different imaging agents and therapeutic drugs into a single nanoparticle (NP) has made it possible to exhibit multiple functionalities [1-4]. Extensive efforts have also allowed for the identification of active targeting reagents, such as folic acid, hyaluronic acid, aptamers, or transferrin, that bind with high specificity to the cancerous cell membrane [5-8]. At the same time, strategies were developed to achieve a controlled drug release triggered by intrinsic physiological microenvironment changes (pH, redox, enzyme, heat, etc.) and/or external stimuli (including light, magnetic/electric field, ultrasound, *etc.*) [9-11]. All those developments taken together, it becomes possible to simultaneously reduce the side-effects of anticancer agents to normal tissues and enhance their therapeutic efficiency [12]. Indeed, there still is some room for large improvement, in terms of drug loading and targeting efficiency.

From an imaging perspective, signals of most previously-reported nanosystems are "always on" regardless of the absence or presence of the target cells, resulting in a low contrast detection [13]. Among available nanoparticles, Mesoporous Silica Nanoparticles (MSNs) have been widely considered for the delivery of anticancer drugs [11,14,15]. To tackle MSNs intrinsic limited loading capacity, Hollow MSNs (HMSNs) with rattle or hollow structure have been prepared as they can efficiently accommodate drugs not only into mesoporous channels of their shell but also within their internal cavity [16-18]. Moreover, recent progress in the design of gated HMSNs has shown some promise in the development of controlled-release theranostic nanosystems. Different "gatekeepers", such as organic molecules, biomacromolecules and nanoparticles, have been used, allowing pore-opening under the stimulation of pH change, temperature, nucleotides, antibodies, enzymes, glucose, or photoirradiation [19-21]. However, these gated nanosystems usually require a highly-sophisticated design that has a strong impact on synthesis time and development cost, ultimately hindering their transfer to the market, despite their demonstrated functionality.

Recently, $MnO_2$ nanosheets, a 2D ultrathin semiconductor material with wide applications in the energy field [22], has attracted large attention as a gatekeeper for drug nanocarriers [23-25]. They exhibit a broad and intense absorption band at around 374 nm, making $MnO_2$ nanosheets an efficient broad-spectrum quencher [26]. Moreover, $MnO_2$ can be converted to $Mn^{2+}$ via reduction by endogenous glutathione (GSH), leading to decomposition of the $MnO_2$ nanosheets [27]. While manganese(IV) ions in the nanosheets are shielded from water due to their 6-fold coordination with oxygen atoms, free solvated Mn(II) ions are efficient $T_1$ contrast agents for Magnetic Resonance Imaging (MRI) [28]. GSH is an essential endogenous antioxidant that has many cellular functions and high GSH levels are implicated in many diseases typically associated with cancer, liver damage, or heart problems [29]. Therefore, much attention has been paid to the use of GSH to design intracellular controlled release systems [30]. In addition, whereas the extracellular pH of normal tissues and blood is constant at 7.4, the measured extracellular values of most solid tumors range from pH 6.5 to 7.2, such lower pH values being more favorable for the reaction between $MnO_2$ and GSH [31]. Based on this principle, several $MnO_2@SiO_2$-based controlled-release nanosystems with GSH-induced contrast-enhanced magnetic resonance signal have already been described [23, 32,33]. However, none of them took advantage of the specificity of HMSNs so far.

Another key challenge in this area lies in the precise control of the drug release using stimuli-responsive systems. Thus, tumor microenvironment, that has a crucial role in cancer progression, exhibit several specific biological and physico-chemical features that can be exploited to trigger the drug release at the tumor site [34]. However, the efficiency and stimulation of nanosystems are often tested in two-dimensional (2D) systems, in which seeded cancer cells do not experience the real microenvironment they find in *vivo*, where physiological fluids, tissues as well as interactions between cancer and stromal cells may impair the drug delivery or its functionality. The transfer from 2D data to *in vivo* during preclinical studies may therefore be extremely time-consuming and costly. Therefore, the design of three-dimensional (3D) environments exhibiting some features of *in vivo* tumors, such as three-dimensional architecture, cell-cell interaction and hypoxia should provide highly useful tumor tissue *in vitro* models for testing anticancer therapeutics [35,36]. In this context, the use of type I collagen, the major protein in most animal tissues, to prepare biomimetic constructs is of particular relevance [37-39]. Most importantly, it has been shown that they could act as models to study the interactions between nanoparticles and cells in a 3D environment [40,41].

Herein, we propose, for the first time, a novel and facile strategy for the fabrication of a multifunctional nanoprobe for contrast-enhanced bimodal cellular imaging and targeted therapy, as depicted in Figure 1. The nanoprobe consists of three components, including (1) doxorubicin(DOX)-loaded hollow mesoporous silica nanoparticles, (2) $MnO_2$ nanosheets that act as both gatekeeper for DOX release from HMSNs and contrast agent for MRI and (3) cancer cell-targeting aptamers (AS1411), that bind to nucleolin, a nucleolar phosphoprotein which is overexpressed on the surface of certain cancer cells [28, 42, 43]. We demonstrate that this multifunctional HMSNs@$MnO_2$(DOX)/apt theranostic nanosystem exhibits a synergistic delivery/imaging effect in standard 2D conditions. We also show that the functionality and specificity of these nanoplatforms are preserved within 3D cellularized collagen hydrogels, that can be useful biomimetic models for evaluating nanotheranostic systems performance before *in vivo* studies.

# Methods

## Materials and reagents

All chemical reagents were analytical grade and used without further purification. Cetyltrimethylammonium bromide (CTAB), tetraethyl orthosilicate (TEOS), sodium carbonate ($Na_2CO_3$), absolute ethanol, concentrated hydrochloric acid, ammonium aqueous solution (25-28%), triethanolamine (TEA), 3-Triethoxysilylpropylamine (APTES), potassium permanganate ($KMnO_4$), 2-(N-Morpholino)ethanesulfonic acid hydrate (MES), doxorubicin hydrochloride (DOX) and fluorescein isothiocyanate (FITC) were purchased from Sigma-Aldrich. MilliQ water (18 MΩ, Millipore, France) was used for the preparation of the solutions and for all rinses. 4,6-diamidino-2-phenylindole (DAPI). Fetal bovine serum, Dulbecco's Modified Eagle's Medium (DMEM), trypsin, glutamine, penicillin-streptomycin solution and AS1411 aptamer (5′-$NH_2$-GGTGGTGGT GGTTGTGGTGGTGGTGG-3′) were purchased from ThermoFisher.

## Preparation of HMSNs@$MnO_2$/apt

Bare HMSNs were prepared using previously reported methods [44,45]. To functionalize the particle surface with amine groups, as-synthesized HMSNs were first dispersed in 20 mL of toluene, followed by addition of 1 mL of APTES. The system was sealed and refluxed at 120 °C in oil bath for 12 h. Afterward, the mixture was centrifuged and washed with ethanol for several times to remove the residual APTES. Then, 20 mg of HMSNs-$NH_2$ was dispersed in 4.2 mL MES buffer (0.1M, pH 6.0) and then 0.8 mL of 5 mM $KMnO_4$ in water was added to the mixture under ultrasonic condition. The resulting mixture was sonicated for another one hour during which brown-black colloids were observed. Subsequently, the raw product was collected by centrifugation, washed several times with deionized water and alcohol to remove any possible residual reactants, and redispersed in 2 mL PBS solution (pH 7.4). The physical adsorption of aptamer on HMSNs@$MnO_2$ was carried out by mixing 0.8 mL of HMSNs@$MnO_2$ (20 mg/mL) in PBS and 200 µL of AS1411 (1 µM). After 3 h of incubation, the solution was centrifuged at 6000 rpm, washed several times, and then dispersed in PBS (pH 7.4) for further application. Full protocol is available in Supplementary Material. The preparation of fluorescently-labeled hollow mesoporous silica nanoparticles was performed by adding FITC to the starting solution used for bare HSMNs preparation.

## Particles characterization

Transmission Electron Microscopy (TEM) studies were performed on a FEI Tecnai G2 Spirit microscope operating at 120 kV. Scanning electron microscopy (SEM) images were obtained on a Hitachi S-3400N scanning electron microscope with a field emission electron gun. Nitrogen sorption-desorption isotherms were measured at 77 K with a Micromeritics ASAP2010 analyzer. Fluorescence images were recorded on a LEICA microscope. Dynamic Light Scattering (DLS) was used to determine the hydrodynamic diameter of the nanoparticles in Milli-Q water or in culture medium with Mastersizer 3000 Particle Size Analyzer. The reading was carried out at an angle of 90° to the incident beam (632 nm). The Contin algorithm was used to analyze the autocorrelation functions. X-ray Photoelectron Spectroscopy (XPS) analyses were performed with a PHOIBOS 100 spectrometer from SPECS GmbH. Inductively Coupled Plasma Mass Spectrometry (ICP-MS) was used to determine the Mn content of the nanoparticles.

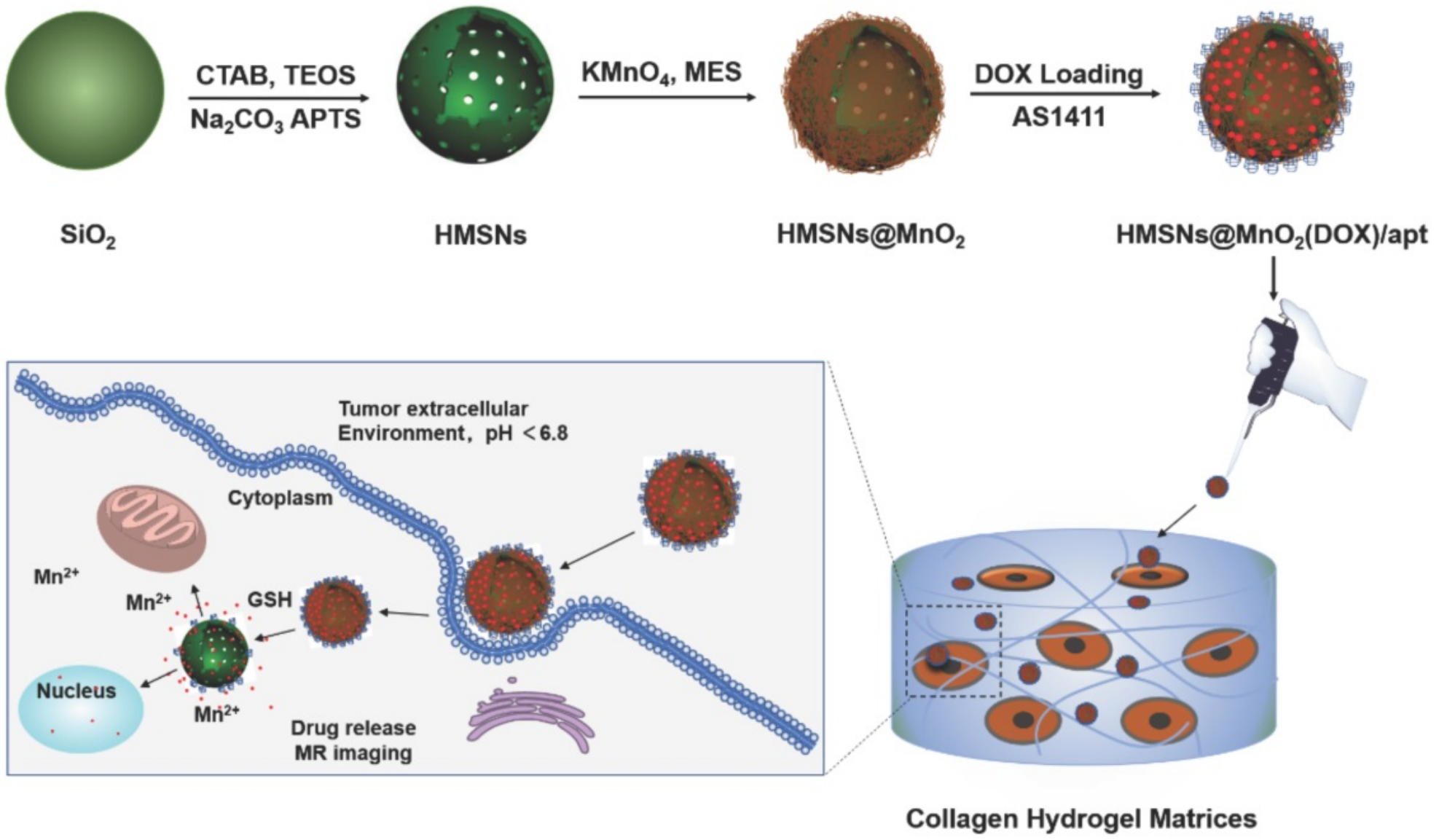

**Figure 1.** Schematic illustration of the preparation of HMSN@$MnO_2$(DOX)/apt and the drug release mechanism.

### Drug loading and release *in vitro*

10 mg of HMSNs@$MnO_2$ nanoparticles were mixed with 1.5 mg of DOX in 1.5 mL of PBS/DMSO (1:1) mixture solution and then stirred under dark conditions for 24 h. Subsequently, the product was collected by centrifugation and washed several times with PBS to remove the free DOX. Then 100 µL of AS1411 (1 µM) in PBS was added, and after 3 h of stirring, the resulting HMSNs@$MnO_2$(DOX)/apt nanoparticles were collected by centrifugation and re-dispersed in PBS solution for subsequent use. To evaluate the DOX loading, the remaining DOX molecules in the supernatant solutions were determined by fluorescence spectroscopy ($\lambda_{ex}$= 500 nm, $\lambda_{em}$ = 590 nm). The loading of the nanoparticles was expressed as the mass percentage of DOX with respect to the total mass of HMSNs@$MnO_2$(DOX)/apt nanoparticles.

*In vitro* DOX release from the HMSNs@$MnO_2$(DOX)/apt nanoparticles was studied in PBS buffer in absence or presence of GSH at pH values of 7.4 and 5.5. For each release study, 1.0 mL of DOX-loaded nanoparticles (1.0 mg.mL$^{-1}$) were dispersed inside a dialysis bag that was soaked in 9.0 mL PBS and shaken at room temperature. At selected time intervals, the sample was collected and 2 mL of solution outside the dialysis bag was removed. Then, 2 mL of fresh PBS buffer was added. The removed solution was properly diluted and the amount of DOX molecules present was measured by fluorescence spectroscopy. The same amount of the DOX solution was introduced in the dialysis bag and its diffusion kinetics was used as control for drug release studies.

### Cell culture

Normal Human Dermal Fibroblasts (NHDF) and HeLa cells were cultured in complete cell culture medium (DMEM with GlutaMAX™, without phenol red supplement, with 10% fetal serum, 100 U.mL$^{-1}$ penicillin, 100 µg.mL$^{-1}$ streptomycin). Tissue culture flasks (75 cm$^2$) were kept at 37 °C in a 95% air: 5% $CO_2$ atmosphere. Before confluence, the cells were removed from culture flasks by treatment with 0.1% trypsin and 0.02% EDTA. Cells were rinsed and resuspended in the above culture medium before use.

### *In vitro* cytotoxicity assays

First, HeLa were seeded in a 24-well plate at a density of ~5×10$^4$ cells/well overnight to allow cell attachment onto the surface of the wells. Then, 0.5 mL of fresh medium containing various concentrations of HMSNs@$MnO_2$(DOX)/apt was added into the wells. After incubation for 24 h, cell activity was evaluated by the Alamar Blue assay. Control experiments were performed by incubating HeLa cells with free DOX and HMSNs@$MnO_2$(DOX) at various drug contents or HMSNs@$MnO_2$(DOX)/apt at equivalent particle concentration for 24 h. To confirm the cytotoxicity of DOX released from the nanoplatform under GSH, similar control experiments were performed with NHDF cells incubated in presence of 5 mM GSH.

### Cell uptake studies by fluorescence imaging

NHDFs and HeLa cells were seeded in a round glass disk that was inserted in 24-well plate (~5×10$^4$ cells per well) and cultured for 24 h. The cell medium was removed, and then cells were incubated with 0.5 mL of fresh cell medium containing FITC-labeled HMSNs@$MnO_2$(DOX)/apt nanoparticles for another 12 h. After medium removal, the cells were washed with PBS for several times, and then 0.5 mL of fresh cell medium with or without 5 mM GSH was added and incubated for another 3 hours. Fixation of the cells was carried out using 4% (v/v) paraformaldehyde (PFA) for 2 h and rinsed three times using PBS. After washing, cellular nuclei were stained with 1% (v/v) solution of DAPI in PBS buffer for 20 min and washed with PBS three times. Last, cell imaging was carried out by fluorescence microscopy. For the HeLa cells imaging, the procedure was similar, except that HMSNs@$MnO_2$(DOX) or HMSNs@$MnO_2$(DOX)/apt particles were added and no GSH treatment was performed.

### Measurement of MRI relaxation properties

Imaging of MRI phantoms measurement was performed with a Bruker Avance III HD 300 spectrometer, equipped with a 10 mm micro imaging probe, having a maximum gradient capacity of 3 T.m$^{-1}$ in the x, y and z directions. The Multi Slice Multi Echo (MSME) pulse sequence was used, acquiring 1 echo but 8 slices with a slice thickness of 1 mm. The size of the images was: 128×128 with a Field of View of 9.5×9.5 mm, resulting in a voxel size of 74×74 µm. Images were acquired with 1 scan and $T_1$ weighting was obtained by using a short repetition time (TR) of 200 ms and using the shortest possible echo time (TE), namely 4.92 ms, in order to minimize the effect of $T_2$ relaxation.

### Preparation of 3D collagen-based tumor models

Type I collagen was purified from rat tails and the final concentration was estimated by hydroxyproline titration, as previously described [37]. Tubes separately filled with collagen solution (2 $mg.mL^{-1}$ in 17 mM acetic acid), whole cell culture medium, and 0.1 M NaOH were kept in ice bathes for 1 h before preparation to slow down the gelling kinetics of collagen. First, 500 µL of collagen solution and 400 µL of culture medium were added to a 1.5 mL tube and vortexed vigorously. After addition of 30 µL of 0.1 M NaOH and strong vortexing, 125 µL of the NHDF or HeLa cell suspension at a density of $10^6$ cells/mL was added and mixed homogeneously. Then 0.9 mL was sampled from the mixture and deposited into a 24-well plate. The plate was then incubated at room temperature for 10 min for complete gelling of collagen.

### Cytotoxicity and functionality within 3D models

Two sets of experiments were designed [41]. In a first approach, selected particles were introduced in the cell suspension prior to their addition to the collagen solution, at a final concentration of 100 $\mu g.mL^{-1}$. Then collagen gels were formed by pH increase to 7. After 48 h of incubation, the cell viability was assessed in the same way as for 2D tests except that 800 µL of water were first added to the collagen gel, left for 0.5 h at room temperature in order to extract the Alamar Blue solution trapped in the gel, and then collected for the absorbance measurements. Particle internalization was also studied in the same configuration. For this, after the 48 h incubation period, the gels were rinsed 3 times with PBS and fixed with 4% paraformaldehyde overnight. Next, the fixed samples were dehydrated in ethanol and butanol, and incorporated in paraffin to be able to obtain 10 µm histological sections with a manual microtome. Before observation, the as-obtained samples were immersed in toluene, ethanol, and then water for rehydration. The cell nuclei were stained with DAPI for 10 min and rinsed with PBS before observation. The second approach consisted in adding 1.0 mL of a 0.2 $mg.mL^{-1}$ suspension of the HMSNs@$MnO_2$(DOX)/apt particles onto the surface of particle-free cellularized collagen gels. After 3 hours of contact, the MR imaging experiments were performed as described above.

### Statistical Analysis

Graphical results are presented as mean ± SD (standard deviation). Statistical significance was assessed using one-way analysis of variance (ANOVA) followed by Tukey (compare all pairs of groups) or Dunnett (compare a control group with other groups) post-hoc test. The level of significance in all statistical analyses was set at a probability of $P < 0.05$.

## Results and discussion

### Preparation and characterization of the nanoplatforms

The HMSN@$MnO_2$/apt nanoparticles were prepared according to the process depicted in Figure 1. In brief, monodisperse solid $SiO_2$ nanoparticles (s$SiO_2$) were first prepared using a modified Stöber method. The prepared s$SiO_2$ were then coated with a CTAB/$SiO_2$ shell via base-catalyzed hydrolysis of TEOS and condensation of silica onto the surface of CTAB pre-coated s$SiO_2$. The resulting s$SiO_2$@CTAB/$SiO_2$ particles were simultaneously treated with $Na_2CO_3$ to remove s$SiO_2$, and $NH_4NO_3$ to remove CTAB, resulting in HMSNs with hollow cores and penetrating pore channels [44]. APTES was grafted on the surface of nanoparticles to get HMSN-$NH_2$. Finally, ultrathin $MnO_2$ nanosheets were formed onto the surface of HMSN-$NH_2$ thanks to the reaction with MES and $KMnO_4$ to obtain HMSN@$MnO_2$. DOX was then loaded inside or on the surface of the nanoparticles by the impregnation scheme described above. To end, AS1411 aptamers were attached on the surface of $MnO_2$ nanosheets thanks to nucleobase-mediated physisorption [28].

As shown in the TEM image of Figure 2a, the prepared hollow mesoporous HMSN nanoparticles exhibit a uniform diameter of ~140 nm and form a colorless well-dispersed suspension in deionized water (DI) (Figure 2d). The diameter measured by DLS, $D_m$, is about 200 nm in DI, which is slightly larger than the TEM data and may reflect some aggregation (Table 1). The contrast between shell and core of the silica nanospheres confirms their hollow structure. The shell of the HMSN with a thickness of ~25 nm displays an obvious mesoporous silica structure generated by the removal of pore templates. In DI water, the obtained HMSNs exhibited a negative zeta potential ζ of -18.4 mV, as expected for silica surfaces (Table 1).

**Table 1.** Mean diameter $D_m$ from DLS with corresponding polydispersity index (PDI) and Zeta potential ζ in deionized water

| Sample | $D_m$ DLS (nm) | PDI | ζ (mV) |
|---|---|---|---|
| HMSNs | 200 ± 18 | 0.254 | -18.4 ± 0.5 |
| HMSNs-$NH_2$ | 256 ± 27 | 0.317 | +10.4 ± 0.3 |
| HMSNs@$MnO_2$ | 319 ± 41 | 0.579 | -15.4 ± 0.2 |
| HMSNs@$MnO_2$(DOX) | 316 ± 26 | 0.500 | -7.3 ± 0.8 |
| HMSNs@$MnO_2$(DOX)/apt | 248 ± 15 | 0.279 | -16.1 ± 0.3 |

The $N_2$ adsorption-desorption isotherm of the HMSN showed a typical type IV curve of surfactant-assisted mesoporous silica with a double strong and sharp adsorption step at intermediate relative partial pressure values around 0.4 (Figure S1). This feature is associated with nitrogen condensation inside the mesopores by capillarity. The application of the Brunauer-Emmett-Teller (BET) model resulted in a high value for the specific surface of 840 $m^2.g^{-1}$. In parallel, the Barrett-Joyner-Halenda (BJH) model applied on the adsorption branch of the isotherms led to an average pore diameter of 3.89 nm (Figure S1). The absence of a hysteresis loop in this interval of $P/P_0$, as well as the narrow BJH pore size distribution, also suggested the existence of uniform cylindrical channels throughout the material.

The successful surface functionalization of the hollow mesoporous silica nanoparticles using APTES could be checked by the increase of the zeta potential that reached a positive value (+10.4 mV) (Table 1). Upon contact of these HMSN-$NH_2$ particles with MES and $KMnO_4$, a visual color change from white to brown was observed (Figure 2e). After extensive washing, UV-vis spectra clearly evidenced a strong absorption band at 400 nm, that can be attributed to the $MnO_2$ nanosheets, whose intensity increase with the initial $KMnO_4$ concentration (Figure S2). The formation of composite nanoparticles (HMSN@$MnO_2$) with a rough capping layer was confirmed by TEM images (Figure 2b). Moreover, the average diameter of HMSN@$MnO_2$ was changed from 200 nm to *ca*. 320 nm, accompanied by a polydispersity index (PDI) increase from 0.254 to 0.579, which suggests that the $MnO_2$ coating induces a slight tendency for the particles to aggregate. The zeta potential of nanoparticles was also modified, from +10.4 mV down to -15.4 mV after functionalization, in agreement with values in the literature [46]. After coating, FTIR spectra also evidenced a peak at 1413 $cm^{-1}$ characteristic of Mn-O vibrations in $MnO_2$ (Figure S3). Sorption isotherms also confirmed the coating of the nanoparticles. HMSN@$MnO_2$ presented flat sorption curves when compared to those of HMSN (at $P / P_0 = 0.4$) (Figure S1), thus indicating a significant pore blocking. Nonetheless, HMSN@$MnO_2$ showed an additional inflection at $P/P_0 > 0.8$, that can correspond to secondary macropores resulting from the $MnO_2$ capping layer with rough surface. HMSN@$MnO_2$ showed a weak peak at pore diameter of 2.35 nm, confirming the partial pore blocking effect of the $MnO_2$ capping layer. Finally, when HMSN@$MnO_2$ were put in contact with GSH (10 mM), the color of the suspension changed back from brown to white, indicative of the disappearance of the $MnO_2$ layer (Figure 2f). TEM images clearly confirmed the regeneration of the smooth surface and clear mesoporous structure of HMSN (Figure 2c).

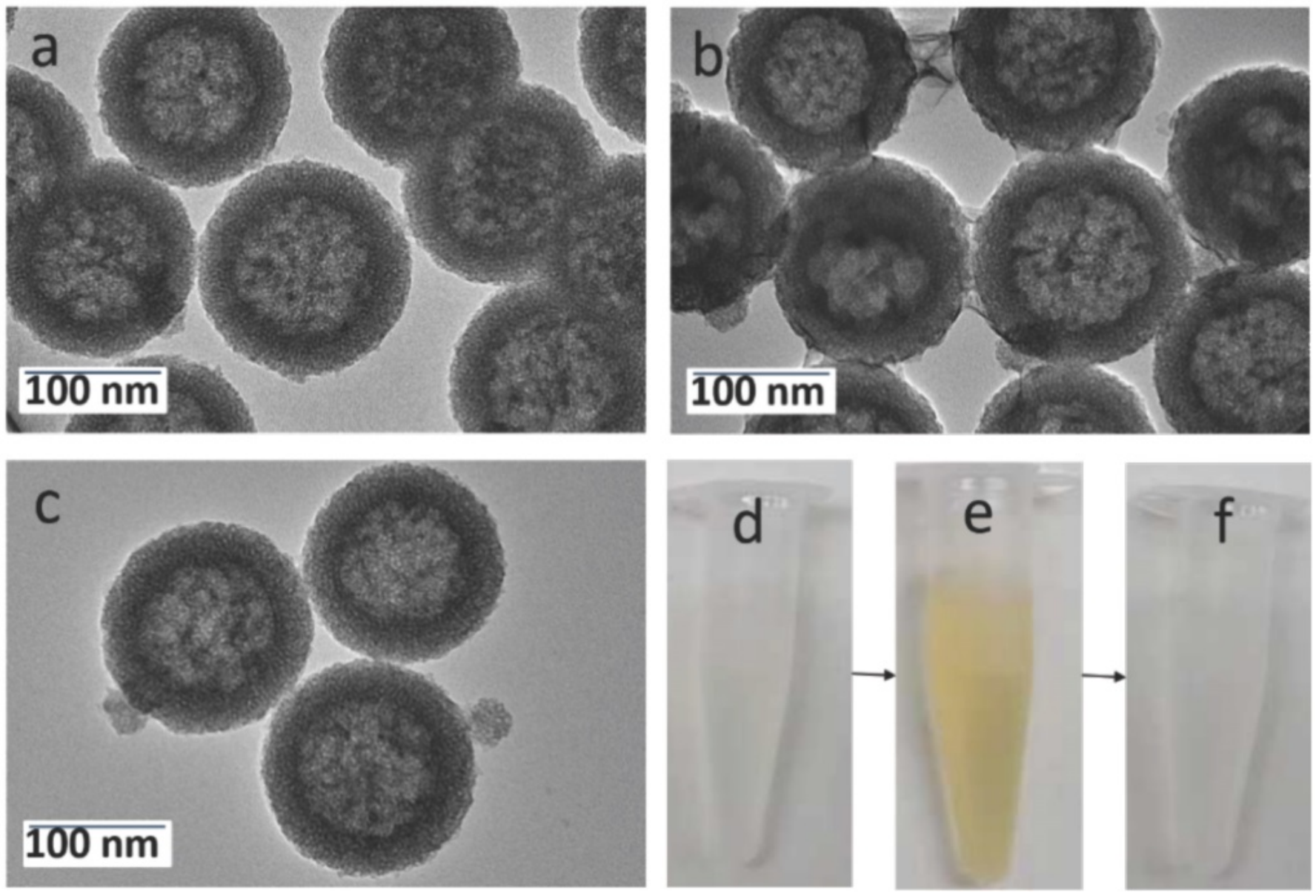

**Figure 2.** TEM images of (a) HMSNs, (b) HMSN@$MnO_2$/apt, (c) HMSN@$MnO_2$/apt after GSH treatment and the corresponding digital images from left to right (d, e, f).

DOX was then selected as a guest molecule to confer anticancer properties to the nanoparticles. The loading was performed by a simple impregnation route in a mixed aqueous/organic medium. The resulting DOX-loaded HMSN@$MnO_2$ particles had a size distribution similar to unloaded particles but the zeta potential changed from -15.4 mV to -7.3 mV, suggesting that at least part of the drug is adsorbed on the $MnO_2$ coating. The UV-vis spectra of the loaded particles showed an additional absorption band at *ca*. 480 nm similar to the one of DOX (Figure S2) and two peaks at 1624 $cm^{-1}$ and 1580 $cm^{-1}$ belonging to the benzene ring structure of DOX could be observed on the FTIR spectra of the HMSNs@$MnO_2$(DOX) sample (Figure S3). The DOX loading of HMSN@$MnO_2$, as estimated by fluorescent emission measurements of the supernatant after impregnation, reached *ca*. 80 $\mu g.mg^{-1}$.

As a final step, the anti-nucleolin AS1411 was adsorbed on the HMSNs@$MnO_2$(DOX) particles which resulted in a significant decrease in the average particle diameter and of the PDI (Table 1). This can be correlated with the more negative value of the zeta potential (from -7.6 mV without aptamer to -16.1 mV after AS1411 adsorption), that results in an enhanced particle colloidal stability thanks to electrostatic repulsion. FTIR spectra showed an additional peak at 1742 $cm^{-1}$ that can correspond to the C=O stretching vibration of the aptamer (Figure S3).

To characterize the chemical compositions of the final HMSNs@$MnO_2$(DOX)/apt nanoparticles, their element composition was determined by XPS (Figure 3a). Mn, Si, C, O and N peaks were observed. Indeed, considering that the analysis depth of XPS analysis is 10 nm at maximum, it only probes the outer surface of the particles. It is therefore not surprising that the Si At% is rather low (19.8 %), corresponding to *ca*. 60 wt% of $SiO_2$. In contrast, the contribution of organic molecules (DOX and AS1411) (> 35 % C) is enhanced compared to the whole particle volume. The Mn relative amount is 0.9 At% corresponding to *ca*. 5 wt % $MnO_2$. Considering that manganese oxide nanosheets have been reported to be *ca*. 1 nm thick, a *ca*. 10 wt % amount would have been expected, suggesting that the $MnO_2$ coating is not homogeneous over the whole particle surface. At the $Mn_{2p}$ level (Figure 3b), the two peaks at 642 eV and 653.8 eV unambiguously sign for the presence of $MnO_2$, as also confirmed by the peak at 530.1 eV on the $O_{1s}$ spectrum [47] (Figure 3d). The peaks at 103.2 eV on the $Si_{2p}$ spectrum (Figure 3c) and 532.4 eV on the $O_{1s}$ spectrum are characteristic of hydrated silica [48]. Deconvoluted spectra at the $C_{1s}$ and $N_{1s}$ levels evidenced the presence of C-C, C-O, C=O, C-N and N-C=O groups (Figure S4). The latter is a clear indication of the presence of the aptamer on the particle surface [49], while the others can belong to both AS1411 and DOX.

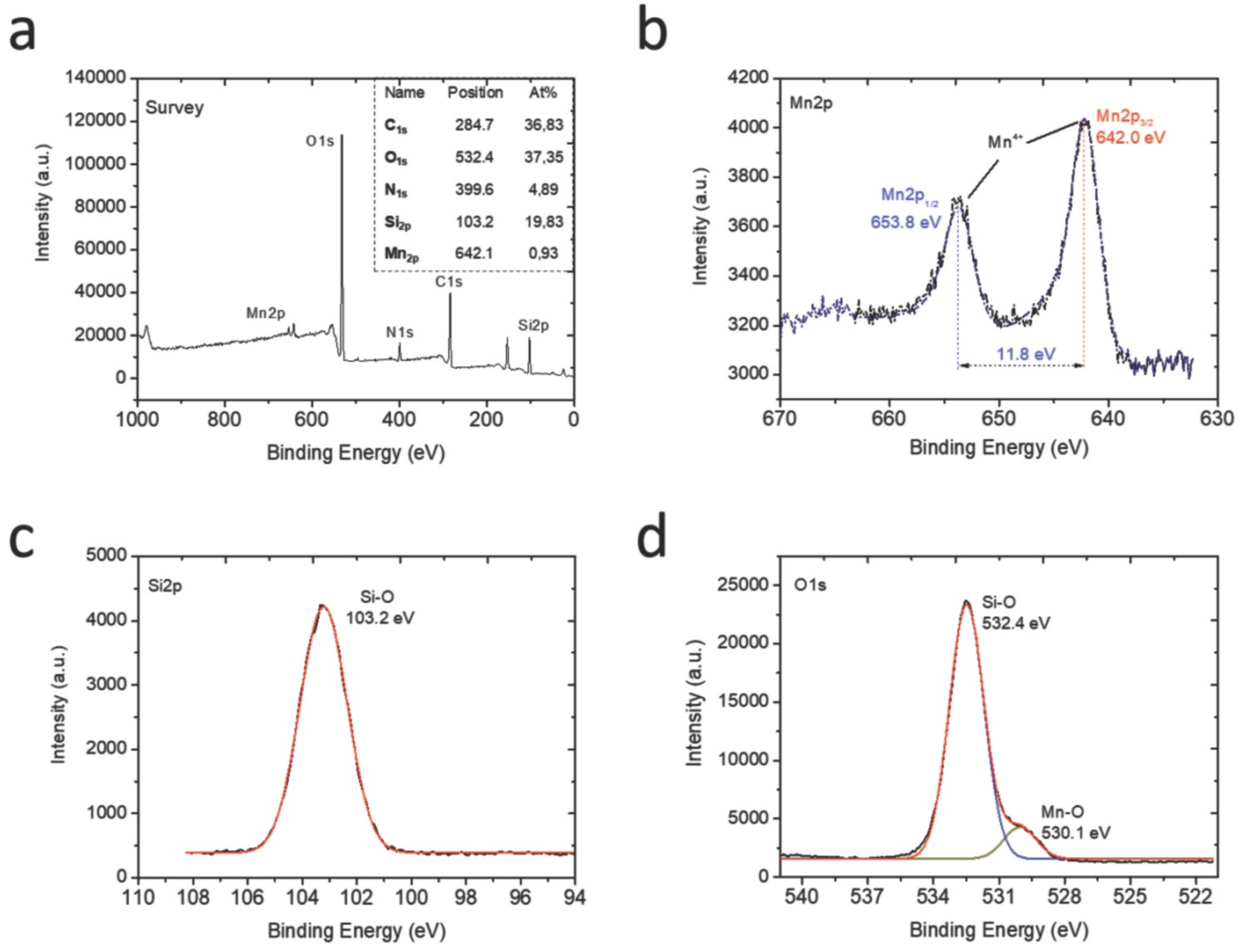

**Figure 3.** XPS analysis of the HMSNs@$MnO_2$(DOX)/apt nanoparticles: (a) full XPS spectrum with atomic analysis and deconvoluted signals with proposed attributions at the (b) $Mn_{2p}$, (c) $Si_{2p}$ and (d) $O_{1s}$ levels.

## Drug release and MRI imaging properties of the nanoplatforms

The feasibility of $MnO_2$ degradation when treated with GSH and pH was first investigated. As shown in the Figure S5, the HMSNs@$MnO_2$/apt nanoplatforms were stable at neutral pH but, after interaction with GSH, the color of the system quickly changed from brown to colorless, with a more pronounced effect when GSH concentration was increased from 5 mM to 10 mM. Acidic pH slightly destabilized the nanoplatform, due to the intrinsic instability of $MnO_2$ in these conditions.

These observations were quantitatively assessed by monitoring DOX release in different conditions (Figure 4). In the absence of GSH, the DOX release was less than 10 % over 50 h at pH 7.4. When the $MnO_2$ coating was destabilized, either in acidic conditions or by addition of GSH at neutral pH, ca. 20 % of the DOX content was released within 10 h and then a slower release, up to ca. 35-40 %, occurred in the next 40 h. When GSH was added in acidic conditions, the shape of the release curve was quite similar (i.e. fast release during 10 h followed by slower release) but the amount of released DOX at the end of the first phase was much higher (ca. 50 %) than in the two previous conditions and the total release after 50 h was larger than 60 %. As pointed out earlier, DOX molecules may be located on the $MnO_2$ surface, inside the porous silica shell and within the particle empty core. Assuming that the GSH-free neutral pH conditions induce no significant dissolution of the $MnO_2$ nanosheets, then the measured low release (less than 10 % of the initial dose) should correspond to surface-adsorbed DOX molecules. Upon GSH addition or acidification, much more DOX molecules are released following a two-step process. It is worth emphasizing that silica nanoparticles are less soluble in acidic than in neutral pH conditions and not sensitive to GSH unless specifically modified [50]. Thus it can be suggested that the first step corresponds to the release of DOX molecules both adsorbed on the surface and trapped in the porous shell due to the $MnO_2$ coating dissolution, while the second one would sign for the slow diffusion of the drug located within the particles. However, when both acidic conditions and GSH addition are combined, the overall release profile, and especially the duration of the fast-release period, is not significantly modified but more DOX is initially released. This may be interpreted considering that, in these conditions, the $MnO_2$ degradation is efficient enough to uncap all pores before diffusion of DOX molecules from the inside of the particle starts. In contrast, in GSH-only or acidic pH-only conditions, the $MnO_2$ degradation process is less efficient and only part of the pores are uncapped before diffusion from the inside compartment becomes effective. This points out that the combination of reductive conditions due to GSH presence and acidic media where $MnO_2$ is less stable is the most favorable for DOX release.

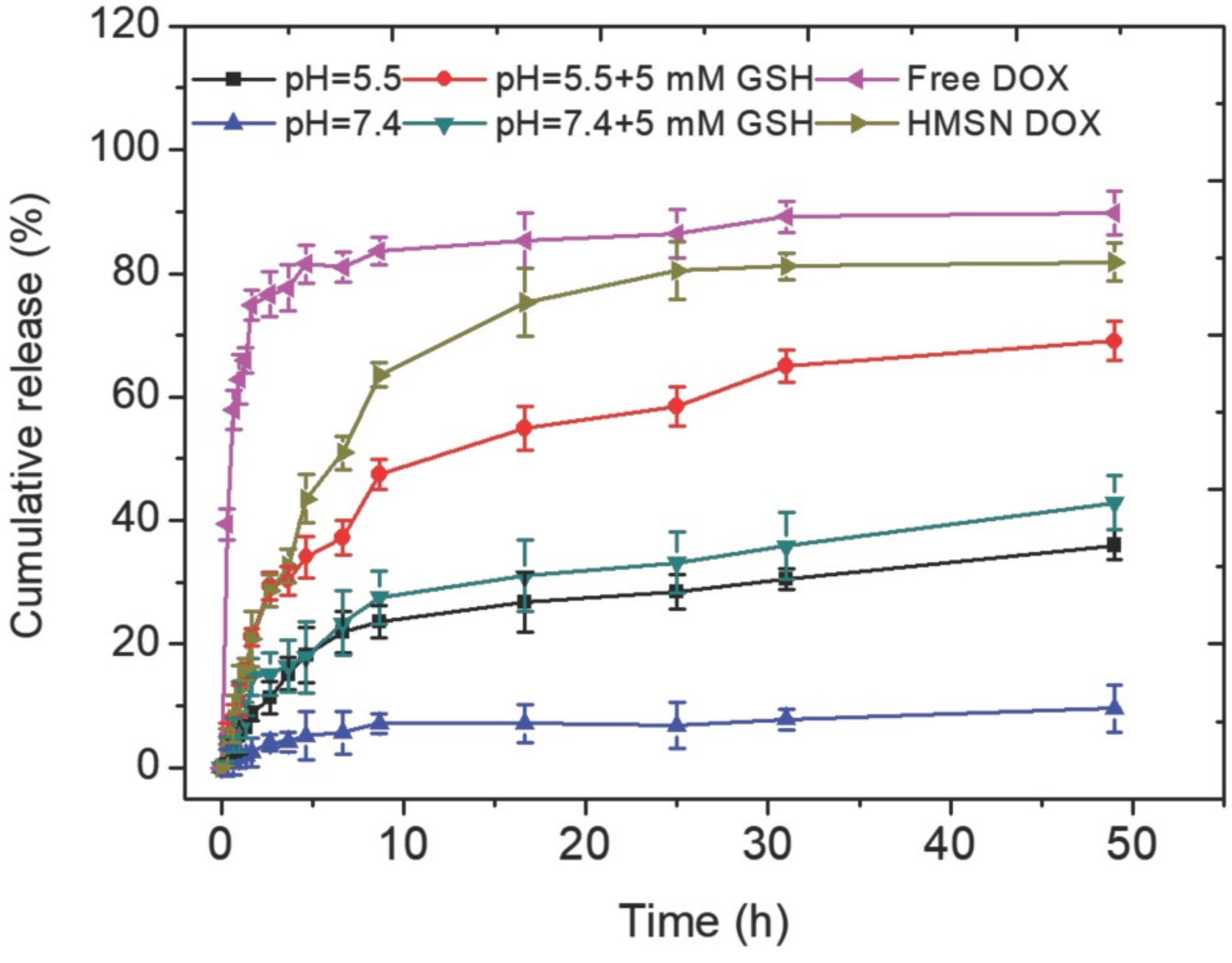

**Figure 4.** Cumulative release of the HMSNs@$MnO_2$(DOX)/apt in different conditions: (blue) pH 7.4, (black) pH 5.5, (green) pH 7.4+ 5 mM GSH, (red) pH 5.5 + 5 mM GSH. Control samples: (pink) free DOX diffusion from dialysis bag and (pale green) uncoated HSMNs(DOX) in PBS solution (pH =7.4).

Because the reduction of $MnO_2$ by intracellular GSH could produce a large amount of $Mn^{2+}$, which are efficient $T_1$ contrast agents, the nanoplatform could also afford GSH/pH-activated detection by MRI. To verify this hypothesis, $T_1$-weighted MR images of HMSN@$MnO_2$/apt aqueous suspensions under different conditions were first performed. As shown in Figure 5a, the MRI signal of the HMSN@$MnO_2$/apt sample in PBS at pH 7.4 didn't exhibit an obvious $T_1$-MRI signal difference compared to that of deionized water. A significant $T_1$-MRI signal enhancement could be observed at pH 5.5. Adding 5 mM GSH in neutral and then acidic conditions further improved the $T_1$-MRI signal intensity. The $r_1$ relaxivity of the nanoplatform suspensions was also measured (Figure 5b) for HMSN@$MnO_2$/apt at pH 5.5 in the presence of GSH, reaching a value of 9.25 $mM^{-1}.s^{-1}$ which is remarkably higher than that of the GSH-free solution at pH 7.4 ($r_1$ = 1.68 $mM^{-1}.s^{-1}$). Importantly such $r_1$ values are higher than those reported for other Mn-based nanocontrast agents ($Mn_3O_4$ nanoparticles, 8.26 $mM^{-1}.s^{-1}$; Mn-MSNs, 2.28 $mM^{-1}.s^{-1}$ and MnO nanoplates, 5.5 $mM^{-1}.s^{-1}$) [51-53]. The above results suggest that the $MnO_2$ nanosheets can efficiently be converted into $Mn^{2+}$ in a mildly acidic/GSH environment, explaining the rapid enhancement of the longitudinal relaxation rate with concentration. In conclusion, HMSNs@$MnO_2$/apt present a clear pH/GSH-responsive $T_1$-MRI performance.

In a second step, it was necessary to check that the designed nanoplatforms could efficiently kill cancer cells while being safe for normal cells. First, it was checked that drug-free HMSNs@$MnO_2$ nanoparticles had no impact on HeLa cells viability after 24 h of contact, as monitored by the Alamar Blue test (Figure S6). Then the cytotoxic effect of HMSNs@$MnO_2$(DOX) and HMSNs@$MnO_2$(DOX)/apt particles on HeLa cells viability was studied with the same method and compared to free DOX at the same drug concentrations. As shown on Figure 6a, no significant difference in cytotoxicity between free and encapsulated drug could be observed, with a continuous decrease of HeLa cells activity with increasing DOX dose, down to *ca*. 50 % for the highest investigated drug dose (16 µg.$mL^{-1}$). The presence of the aptamer had no influence either. In a step further, the effect of HMSNs@$MnO_2$(DOX)/apt particles concentration on HeLa and NHDF cells was monitored (Figure 6b). In the 5-160 µg.$mL^{-1}$ range, HMSNs@$MnO_2$(DOX)/apt particles have no obvious impact on NHDF while they are toxic for HeLa cells (*i.e.* viability below 80 %) for particle concentrations of 10 µg.$mL^{-1}$ or more. Importantly, if GSH (5 mM) was present in the NHDF culture medium, then high cytotoxicity was also measured, confirming its ability to trigger DOX release.

To confirm this hypothesis, the internalization of the HMSNs@$MnO_2$(DOX)/apt was studied. As shown in Figure 7a-d, accumulation of the nanoplatforms, visualized by the green fluorescence of encapsulated FITC, in contact or within the cells could be observed for NHDF, without or with GSH, and HeLa cells. In the latter case, bright green aggregates are located near the nuclei, which appear blue after DAPI staining, strongly supporting their intracellular localization. For this sample, colocalization of the nanoplatform (green signal) and the DOX molecules (red signal) is evidenced. Noticeably, when no aptamer was present on the surface, low particle accumulation was observed and no clear red signal could be imaged (Figure 7d). For NHDF cells without GSH, the red areas are hardly seen whereas they are more clearly visible/detectable when GSH was present. It is important to point out that $MnO_2$ nanosheets have previously been shown to efficiently quench the fluorescence of dyes in the UV-visible range [54]. Thus the observation of red signals on the fluorescence images indicates that DOX is no longer interacting with the manganese oxide coating, probably because the $MnO_2$ nanosheets were reduced to $Mn^{2+}$ by GSH. Noticeably, in the case of HMSNs@$MnO_2$(DOX)/apt interacting with HeLa, the strong co-localization of FITC and DOX inside the cells would suggest that the $MnO_2$ layer has been degraded but that part of the drug remains inside the particles (Figure 7c). In sharp contrast, for HMSNs@$MnO_2$(DOX)/apt interacting with NHDF in the presence of GSH, many particles visualized by monitoring the green fluorescence do not exhibit a red fluorescence (Figure 7b). This can be attributed to the fact that the $MnO_2$ layer is degraded by GSH leading to DOX release before the internalization process. While viability tests have evidenced NHDF cells death in these conditions (blue bars on Figure 6b), indicating that released drug could be uptaken, the dilution of DOX within the cytoplasm may lead to a low fluorescence intensity, precluding its detection.

Altogether, HMSNs@$MnO_2$(DOX)/apt nanoplatforms have the ability to be internalized by HeLa cells with a clear benefit of the presence of the targeting aptamer on the efficiency of this process. After degradation of the $MnO_2$ coating, that may be due to both acidic pH of the lysosome and enhanced GSH expression, the particles can thereby release DOX molecules at the vicinity of the nucleus, resulting in an efficient cancer cell killing. At the same time, the formation of $Mn^{2+}$ ions should allow for the detection of the cancer tissue via $T_1$-weighted MRI. Very importantly, our experiments also indicate that these nanoplatforms are safe for normal NHDF cells in the absence of GSH.

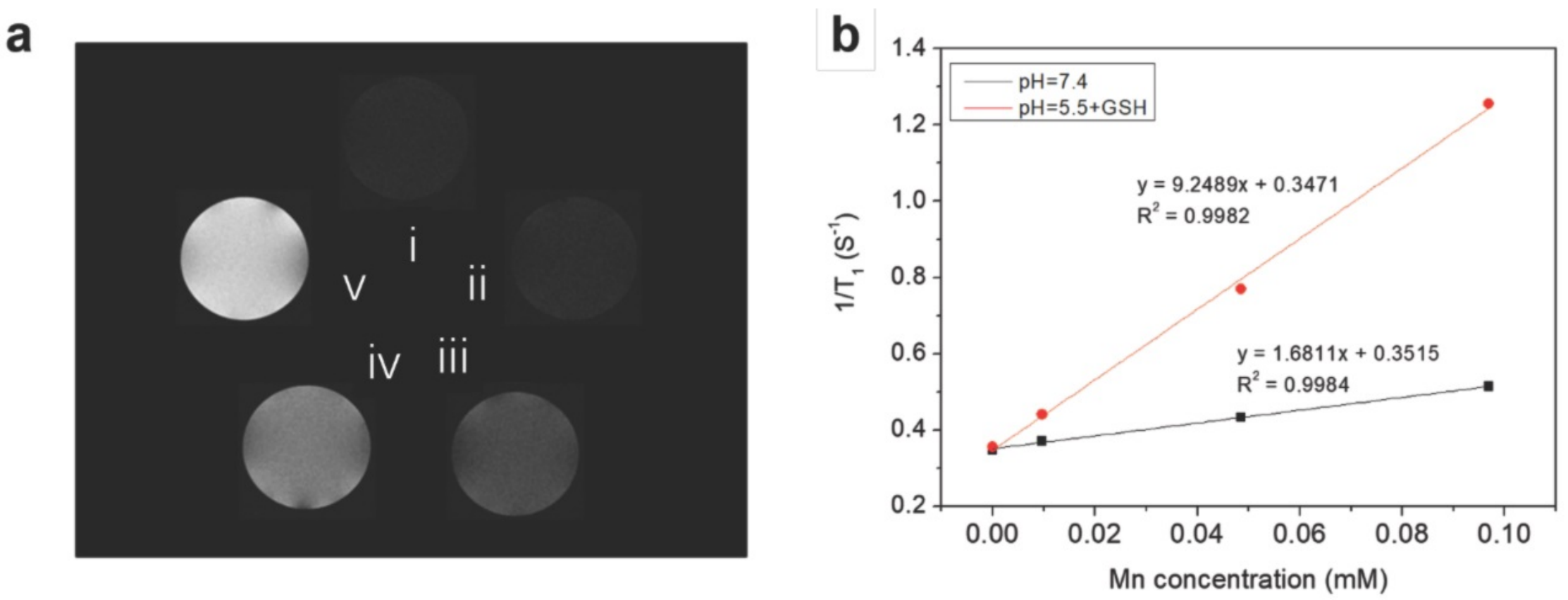

**Figure 5.** (a) $T_1$-weighted solution MR images of HMSN@$MnO_2$/apt under different conditions. (i: $H_2O$; ii: pH 7.4; iii: pH 5.5; iv: pH 7.4 + GSH 5 mM and v: pH 5.5 + GSH 5 mM) after incubation at 37 °C for 1 h. (b) $1/T_1$ versus Mn concentrations for HMSN@$MnO_2$/apt at (black) pH 7.4 and (red) pH 5.5 + 5 mM GSH.

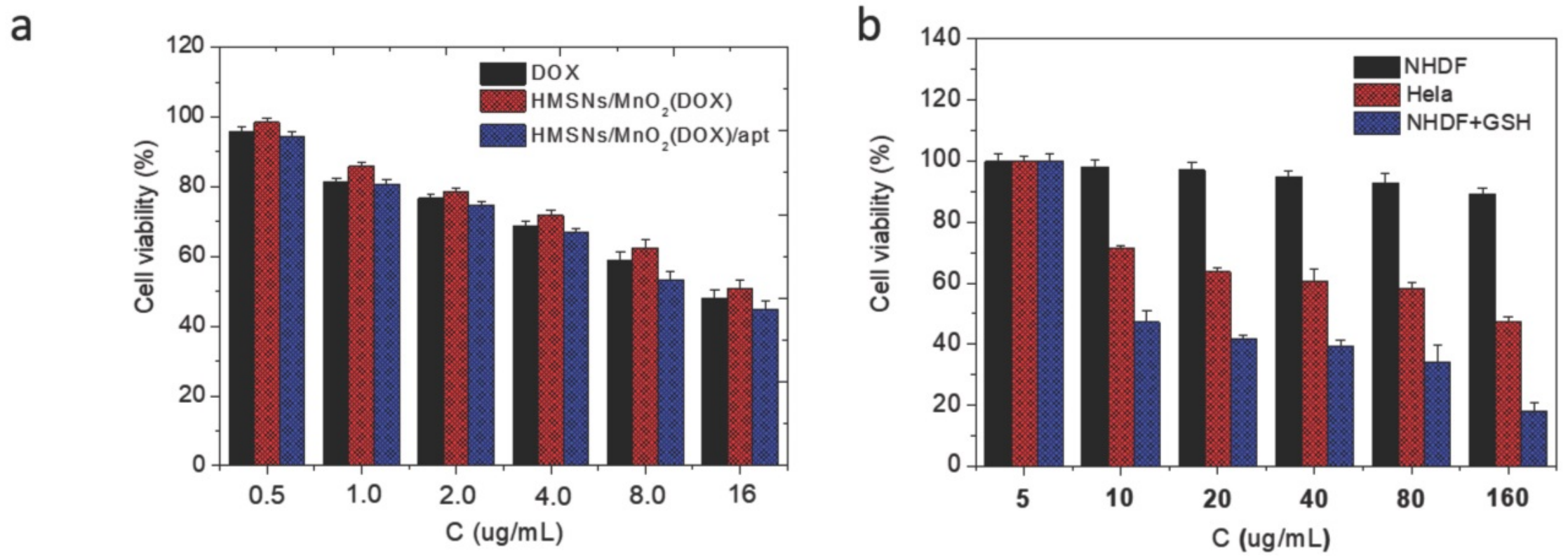

**Figure 6.** (a) HeLa cell viability after incubation with DOX, HMSNs@$MnO_2$(DOX) and HMSNs@$MnO_2$(DOX)/apt for 24 h as a function of DOX dose. (b) Cell cytotoxicity of HMSNs@$MnO_2$(DOX)/apt after 24 h of incubation with NHDF, HeLa and NHDF cells treated with GSH as a function of particle concentration.

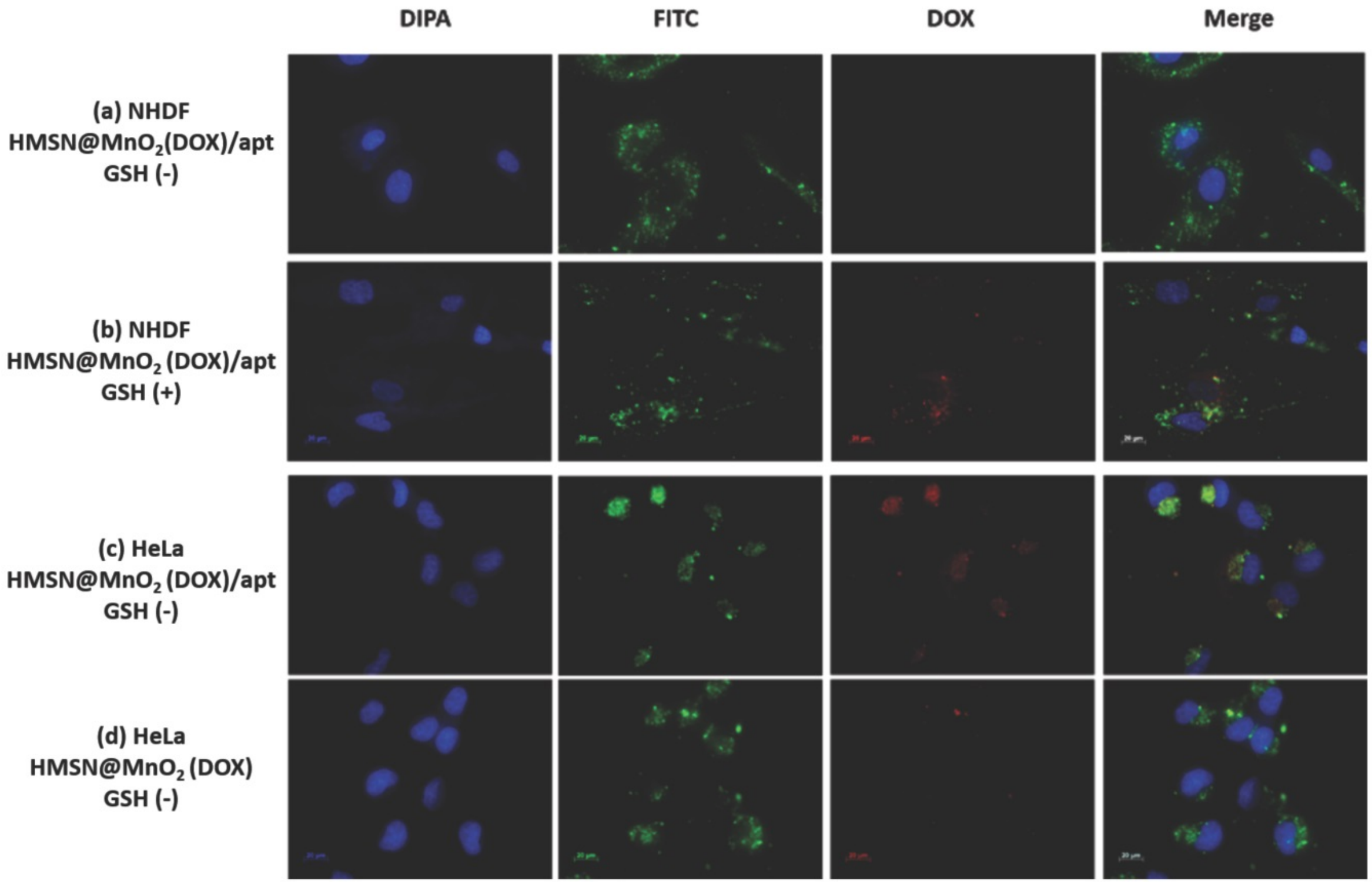

**Figure 7.** Fluorescent images of (a) NHDF cells incubated with HMSNs@$MnO_2$(DOX)/apt nanoparticles, (b) NHDF cells incubated with HMSNs@$MnO_2$(DOX)/apt nanoparticles and 5 mM GSH, (c) HeLa cells incubated with HMSNs@$MnO_2$(DOX)/apt nanoparticles, (d) HeLa cells incubated with HMSNs@$MnO_2$(DOX) nanoparticles.

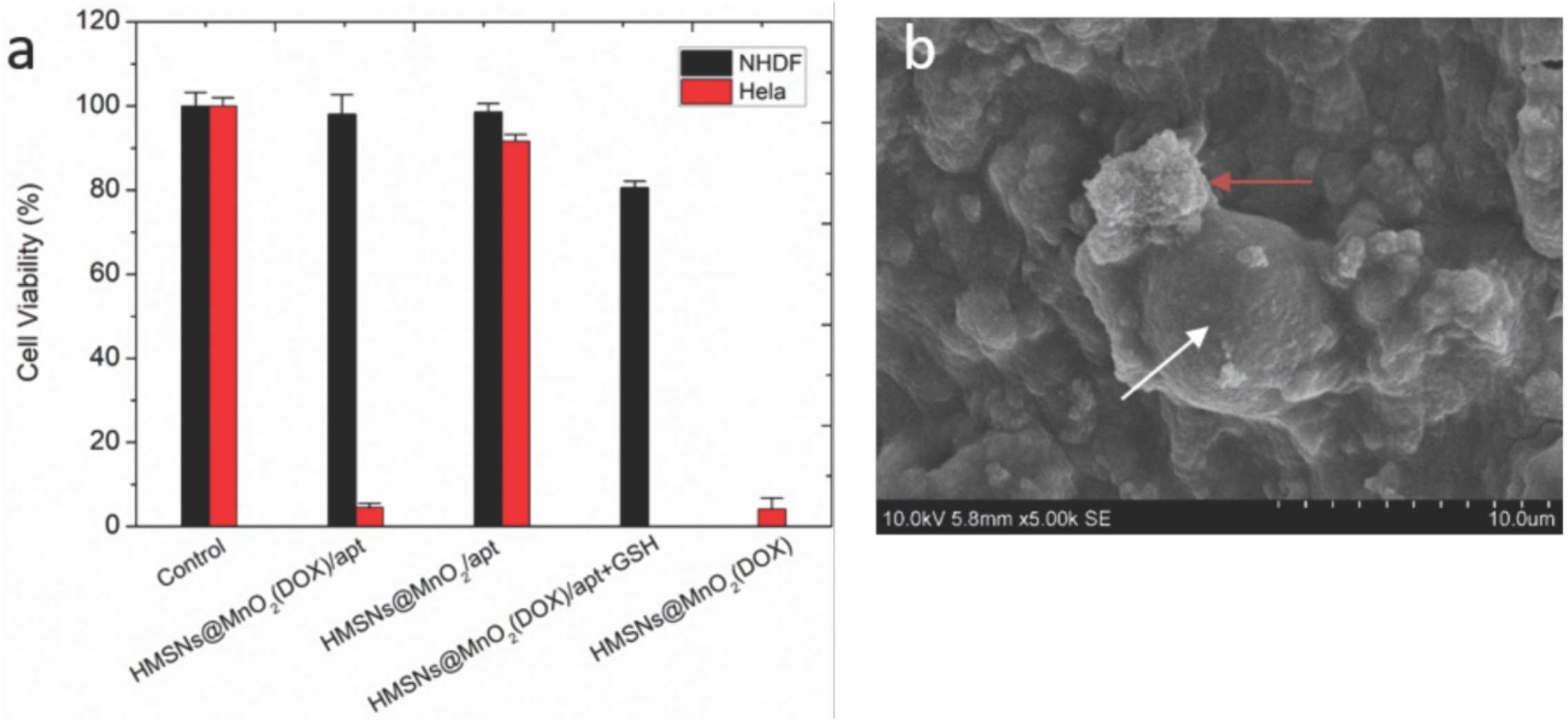

**Figure 8.** (a) Viability of NHDF and HeLa cells within different types of 3D collagen/silica nanocomposites after 24 h of incubation. (b) SEM image of nanoparticles (red arrow) interacting with HeLa cells (white arrow) within collagen hydrogels.

## Functionality of nanoplatforms in 3D biomimetic collagen matrices

At this stage of similar studies, *in vivo* evaluations are usually undertaken using tumor-bearing animals. However, in this work, we chose instead to go deeper into the understanding of the impact of a 3D environment on the functionality of the nanoplatform, a point that was not previously assessed in the literature. For this we prepared type I collagen matrices in conditions compatible with cell immobilization and particle encapsulation or diffusion [40,41].

To check whether the nanoparticles preserved their efficiency and selectivity as cancer cell killing agents, the viability of NHDF and HeLa cells co-encapsulated with HMSNs@$MnO_2$, HMSNs@$MnO_2$(DOX), and HMSNs@$MnO_2$(DOX)/apt at a dose of 100 µg.mL$^{-1}$ within the collagen matrices was studied (Figure 8a). After 24 h, compared to the control group, the unloaded HMSNs@$MnO_2$/apt showed very little toxicity for both types of cells. After DOX-loading, the HMSNs@$MnO_2$(DOX)/apt nanoparticles neither showed much effect the viability of NHDF cells but exhibited a clear cytotoxic effect on HeLa cells, that was independent of the presence of the aptamer on the surface. Importantly, the decrease of HeLa cell viability was more significant than in the 2D experiments (see Figure 6b above), suggesting that the confinement of the cells and particles within the matrices strongly favors their interaction. As a matter of fact, SEM imaging allowed for the observation of nanoparticle aggregates at the close vicinity of the HeLa cells (Figure 8b). In such conditions, our results suggest that it may be no longer necessary to specifically target the cells by using the AS1411 aptamer.

Fluorescence imaging of the encapsulated HeLa cells also evidenced the co-localization of DOX with the nuclei-staining DAPI, signing for their successful internalization (Figure 9). However red fluorescence signals could also be visualized outside the cells, suggesting that some $MnO_2$ degradation also occurred outside the cells.

Then, to evaluate whether such an ultrasensitive GSH-responsive is suitable for *in vivo* tumor imaging by magnetic resonance, another configuration was used where selected nanoparticles were placed onto cellularized collagen hydrogels and left to diffuse for 3 h. As shown in Figure 10, a weak MRI signal was obtained for all samples prepared in the absence of HeLa cells or in the presence of NHDF cells, in agreement with the fluorescence image in Figure 9 that suggested partial destabilization of the nanoplatforms in contact with the collagen network. While comparable results were obtained when the aptamer-free particles were added to HeLa cells, HMSNs@$MnO_2$(DOX)/apt particles led to a significant enhancement of the intensity of $T_1$-weighted images, that was increased when the initial particle concentration was increased. These results clearly demonstrate that the functionality and specificity of the HMSNs@$MnO_2$(DOX)/apt nanoplatforms is preserved within the 3D models. Interestingly, in this configuration, particles are not initially confined at the proximity of the cells but must find their way through the collagen network and bind to the HeLa cells. This explains why, in contrast to the previous system, the presence of the aptamer does have a significant beneficial effect on the MRI signal. In these 3D models, determination of the $r_1$ value is difficult as it would require to determine the precise amount of HMSNs@$MnO_2$(DOX)/apt that have diffused inside the matrices and interacted with the cells. However, when comparing Figure 5a and Figure 10, it quite clear that the $T_1$ relaxation time is lower in the matrix than in suspension and that the decreased stability of the nanoplatforms in the collagen gels ultimately impacts on its contrasting efficacy, i.e. its ability to differentiate safe and tumor tissues.

## Conclusion

As a conclusion, a multicomponent hybrid nanoplatform allowing for combined MRI contrast-enhanced imaging and targeted therapy through controlled drug delivery was designed, prepared and fully characterized *in vitro* in 2D and 3D configurations. Of particular importance is the fact that their high selectivity towards HeLa cancer cells compared to normal human cells was preserved when they were able to diffuse through cellularized biomimetic collagen hydrogels. This strongly suggests that such hydrogels, that are (i) simple to prepare, (ii) suitable for traditional cell biology studies, (iii) amenable to MRI measurements and (iv) can easily be implemented in terms of structure, composition and type of immobilized cells, should be useful for a preliminary screening of newly-developed nanotheranostics before entering *in vivo* studies that are time-consuming, costly and are raising increasing ethical and public concern.

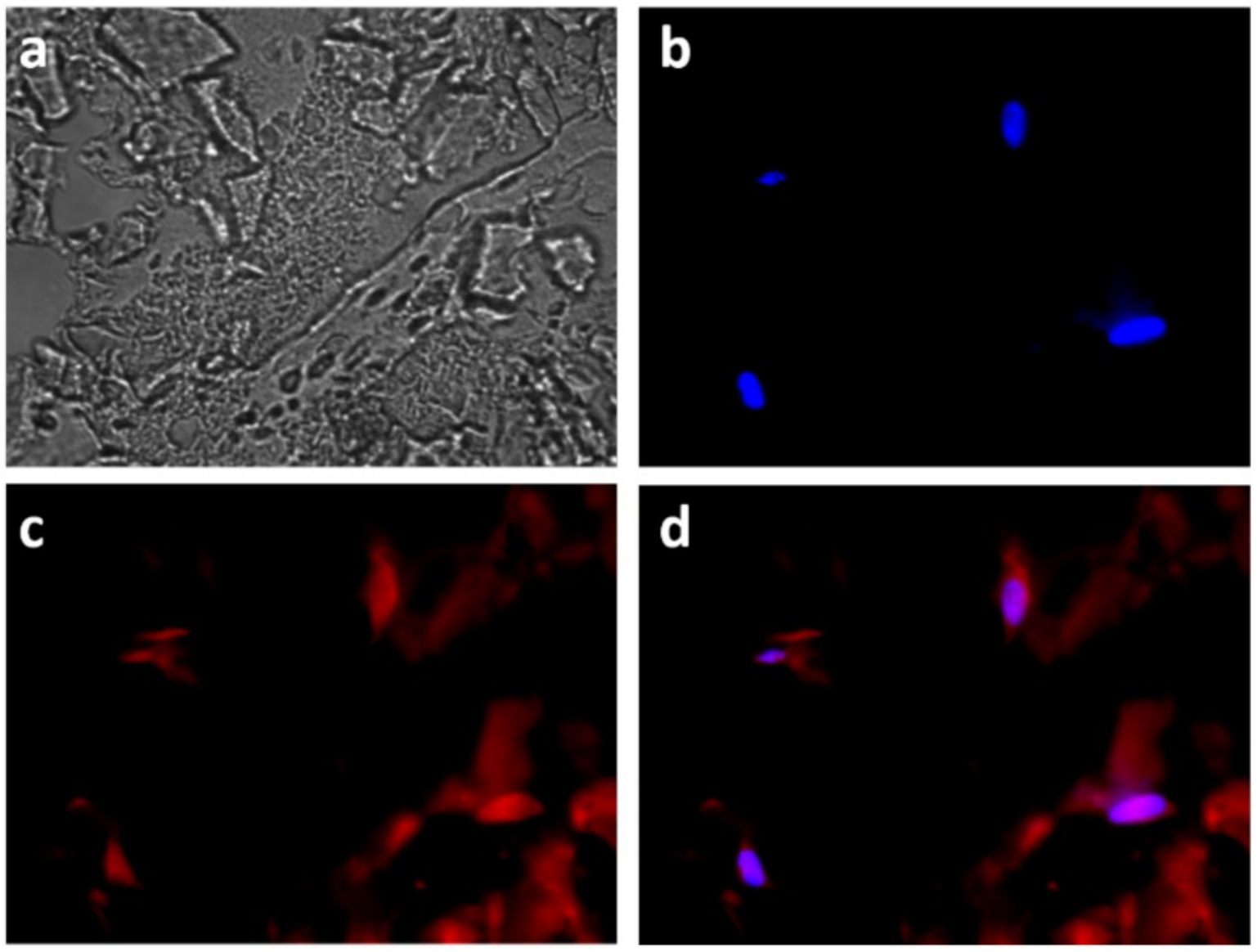

**Figure 9.** Fluorescent images of HeLa cells incubated with HMSNs@$MnO_2$(DOX)/apt within a collagen hydrogel for 24 h, (a) bright field; (b) blue channel (DAPI); (c) red channel (DOX) and (d) merge.

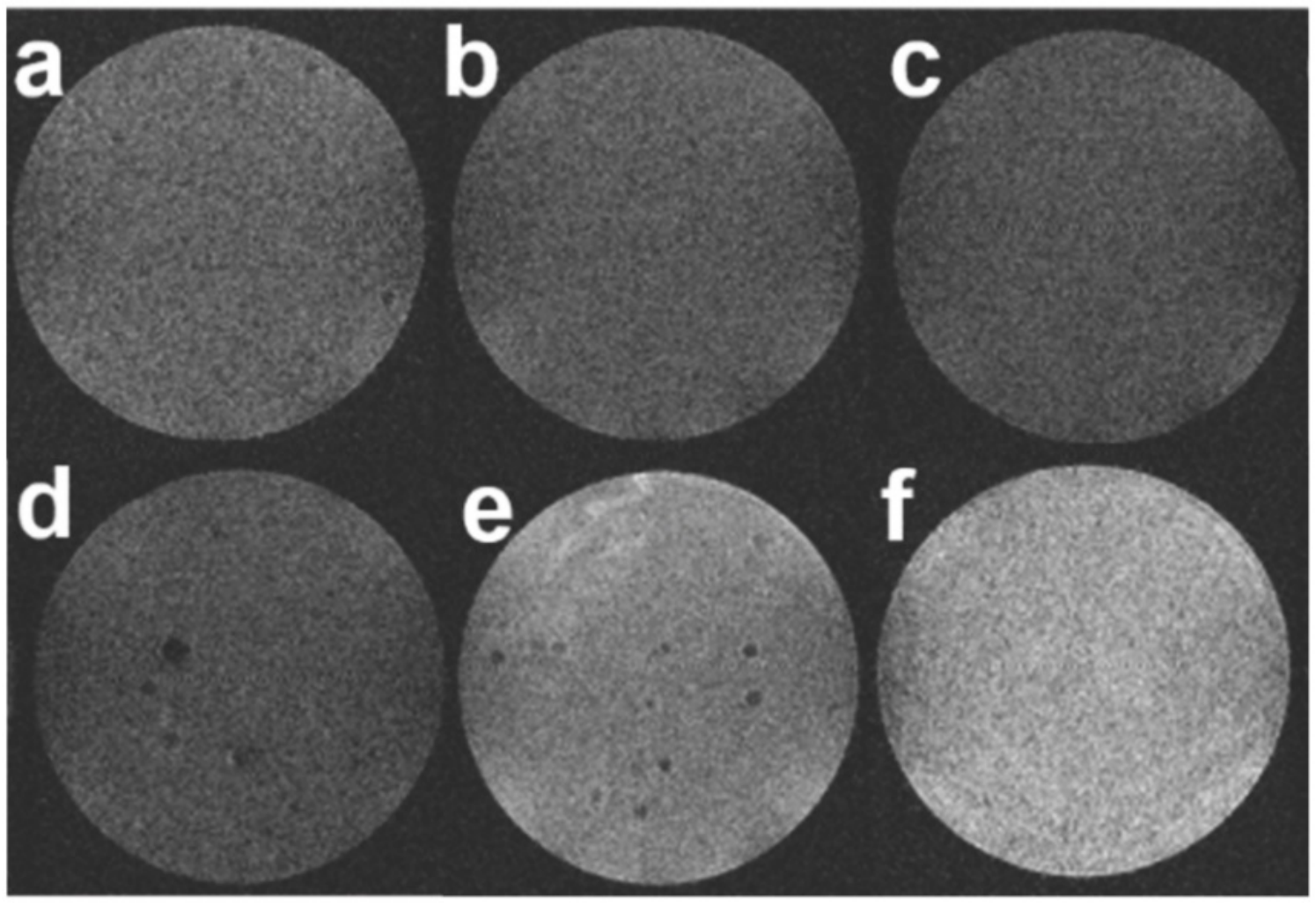

**Figure 10.** $T_1$-weighted MR images of collagen gels (a) without cells, (b,c) with NHDF cells and (d-f) with HeLa cells after 3h diffusion of (b,d) 0.2 mg.mL$^{-1}$ HMSNs@$MnO_2$(DOX) (c,e) 0.2 mg.mL$^{-1}$ HMSNs@$MnO_2$(DOX)/apt and (f) 0.4 mg.mL$^{-1}$ HMSNs@$MnO_2$(DOX)/apt.

## Acknowledgment

Y.S. PhD grant was funded by the China Scholarship Council.

## Supplementary Material

Full protocol for particle synthesis and supplementary figures. http://www.ntno.org/v02p0403s1.pdf

## Competing Interests

The authors have declared that no competing interest exists.